\documentstyle[aps,prb,preprint]{revtex}
\begin{document}
\bibliographystyle{aip}
\draft
\title{Sensitivity of the Superconducting 
Transition Temperature to Changes in 
the Spin-Fluctuation Spectral Weight}
\author{P. McHale\footnote{Present address: Max-Planck-Institut f\"{u}r
Physik komplexer Systeme, N\"{o}thnitzer Str. 38, 01187 Dresden, Germany} 
and P. Monthoux}
\address{Cavendish Laboratory, University of Cambridge
\\Madingley Road, Cambridge CB3 0HE, United Kingdom}
\date{\today}
\maketitle
\begin{abstract}

In the simplest model of magnetic pairing, the transition temperature
to the superconducting state depends on the dynamical susceptibility 
$\chi({\bf q},\omega)$. We discuss how $T_c$ is affected by different
momentum and frequency parts of $\chi({\bf q},\omega)$ for nearly 
antiferromagnetic and nearly ferromagnetic metals in two dimensions.
While in the case of phonon-mediated superconductivity any addition 
of spectral weight to $\alpha^2F(\omega)$ at $\omega >0$ leads to an 
increase in $T_c$, we find that adding magnetic spectral weight at
any momentum ${\bf q}$ and low frequencies  ($[0:3T_c]$ and $[0:(5-9)T_c]$ 
for nearly antiferromagnetic and ferromagnetic metals respectively)
leads to a suppression of $T_c$. The most effective frequency and 
momentum range consists of large momenta ${\bf q} \sim (\pi,\pi)$
and frequencies around $10T_c$ for nearly antiferromagnetic metals
and small momenta ${\bf q} \sim 0$ and frequencies of 
approximately $(13-22)T_c$ for nearly ferromagnetic metals.
\end{abstract}
\narrowtext
\section{Introduction}
\noindent 
In principle there are over one million ternary and over one hundred
million quaternary crystalline materials. Even the binary compounds
number well over eleven thousand. Only a tiny percentage of these
materials have been synthesised. Clearly an exhaustive search of the
periodic table for superconductivity is out of the question. However,
there exist in nature a wide range of compounds which are very close
to a magnetic instability, and in quite a number of cases, an anisotropic
superconducting phase is experimentally found on the border of magnetic
order. But this type of superconductivity is sometimes confined to very 
small regions of the phase diagram, making it difficult to detect. The
heavy-fermion superconductor UGe$_2$\cite{Saxena} is a case in point. An
intuitive understanding of the trends in $T_c$ not only provides a
test of our models of anisotropic superconductivity but can also
guide the experimental search for new anisotropic superconductors.

One of the most extensively investigated models of anisotropic
superconductivity is based on a magnetic interaction, in which 
quasiparticles interact via the exchange of ferromagnetic or 
antiferromagnetic spin-fluctuations. An understanding of the properties
of this model is essential in assessing its ability to describe metallic
systems on the border of magnetic long-range order. The role of 
dimensionality, whether one is close to a ferromagnetic or 
antiferromagnetic instability and the sensitivity of $T_c$ to model 
parameters have been investigated within a mean-field 
framework.\cite{pvsdMon99,3DMon01,Arita00} In these calculations, 
the transition temperature reflects an integral property of the magnetic 
excitation spectrum and does not tell us how much different parts 
of $\chi({\bf q},\omega)$ contribute to the answer.

For the traditional isotropic superconductors, the transition 
temperature $T_c$ depends upon the spectral function $\alpha^2F(\omega)$ 
which characterises the phonon-mediated electron-electron interaction. 
The sensitivity of the transition temperature to variations in the 
spectral distribution  $\alpha^2F(\omega)$ was first investigated
by Bergmann and Rainer\cite{br73}. They found that all frequency regions 
of $\alpha^2F(\omega)$ yield a positive contribution to $T_c$. Frequencies
$\omega \gg T_c$ and $\omega \ll T_c$ contribute little (nothing in 
the limit $\omega \rightarrow 0$) while frequencies around $2\pi T_c$ 
contribute the most.

Later Millis, Sachdev and Varma\cite{MillisSachdevVarma88} extended the 
Bergmann-Rainer analysis of the \'{E}liashberg equations to 
the case of magnetically mediated singlet superconductivity.
They assumed that the dynamical magnetic susceptibility could be 
expressed as a momentum-dependent factor times a frequency-dependent 
factor and that the spin fluctuations could be described by an 
Einstein model. In their model the sensitivity of $T_c$ to changes in
the magnetic spectral weight did not depend on the paramagnon momentum
${\bf q}$ and made the computational analysis of the problem very similar
to that of the phonon case. They showed that, in contrast to the latter
problem, there is a crossover frequency $\omega_{cross}$ such that adding
magnetic spectral weight at frequencies  $\omega < \omega_{cross}$ led
to a reduction in $T_c$.

More recently, Monthoux and Scalapino\cite{ms94}
carried out this analysis for the fluctuation-exchange approximation 
to the two-dimensional Hubbard model. For this more realistic magnetic 
pairing interaction, which is very non-local in space, they
generalized the Bergmann-Rainer approach in order to study how sensitive
the transition temperature to the $d_{x^2-y^2}$ superconducting state is
to infinitesimal changes in the spectral weight at frequency $\omega$ and
momentum ${\bf q}$. In agreement with the results of Millis et 
al\cite{MillisSachdevVarma88}, 
they found that adding spectral weight at very low 
frequencies and for any momentum ${\bf q}$ led to a reduction in $T_c$.
They also found that there was a region in momentum space near 
${\bf q} = 0$ where any small addition of spectral weight at any 
frequency also led to a reduction in $T_c$. They mapped out the region in
${\bf q}$ and $\omega$ space that gave a positive contribution to $T_c$,
essentially wavevectors near ${\bf q} = (\pi,\pi)$ and 
frequencies $\omega > T_c$.

Here we report the results of a similar analysis using a parametrization
of the effective interaction arising from the exchange of magnetic 
fluctuations. We consider systems on the border of antiferromagnetism and
ferromagnetism. We study changes in $T_c$ brought about by small 
variations in the spectral weight at some wavevector ${\bf q}$ and 
frequency $\omega$. We map out the regions of wavevector and frequency space 
where a small addition of spin-fluctuation spectral weight yields an
enhancement or suppression of $T_c$. This will give a more detailed understanding 
of the trends in $T_c$ that one calculates from the model. We shall contrast
these findings with those obtained for the conventional phonon-induced pairing
mechanism\cite{br73} to gain insights into the similarities and 
differences between phonon-mediated and magnetically mediated
superconductivity.
\section{Model}
\noindent
We consider quasiparticles on a two-dimensional square lattice and
postulate the following effective action
\begin{eqnarray}
S_{eff} & = & \sum_{{\bf p}\alpha} \int_0^\beta \mathrm{d}\tau \,
	\psi_{{\bf p}\alpha}^\dagger(\tau) 
	\Big( \partial_\tau+\epsilon_{\bf p}-\mu \Big) 
	\psi_{{\bf p}\alpha}(\tau) \nonumber \\
	&   & \mbox{} - \frac{g^2}{6N} \sum_{\bf q} 
	\int_0^\beta \mathrm{d}\tau
	\int_0^\beta \mathrm{d}\tau' \, \chi({\bf q},\tau-\tau')
	{\bf s}({\bf q},\tau) \cdot {\bf s}(-{\bf q},\tau'),
\label{eq:effac}
\end{eqnarray}
where $\psi^\dagger_{{\bf p}\sigma}$ and $\psi_{{\bf p}\sigma}$ are 
Grassman variables and $N$ is the total number of allowed wavevectors in the
Brillouin Zone. The spin density is given by
\begin{equation}
{\bf s}({\bf q},\tau) \equiv 
\frac{1}{N}
\sum_{{\bf p}\alpha\gamma}
	\psi_{{\bf p+q}\alpha}^\dagger(\tau) 
	\mbox{\boldmath $\sigma$}_{\alpha\gamma}
	{\bf \psi}_{{\bf p}\gamma}(\tau) ,
\end{equation}
where {\boldmath $\sigma$} is the vector whose components are the three
Pauli spin matrices. 

The dispersion relation is  
\begin{equation}
\epsilon_{\bf p} = -2t( \cos (p_x) + \cos (p_y) ) - 4t' \cos (p_x) 
\cos (p_y), \qquad t' \leq 0.5t,
\label{eq:dispersion}
\end{equation}
where for simplicity we use units in which the lattice spacing is unity. 
For comparison with earlier work on this model\cite{pvsdMon99,3DMon01} 
we set $t'=0.45t$ and adopt the value $n = 1.1$ for the band filling.

We parametrize the retarded generalized magnetic susceptibility as:
\begin{equation}
\chi({\bf q},\omega) = \frac{\chi_0\kappa_0}
	{\kappa^2+\hat{q}^2-i[\omega/\eta(\hat{q})]}.
\label{chiML}
\end{equation}

\noindent where $\kappa$ and $\kappa_0$ are the inverse correlation lengths 
with and without strong magnetic correlations respectively. The 
correlation length is related to the pressure applied to the sample with 
$\kappa^2 = 0$ coinciding with the quantum critical point.
Let
\begin{equation}
\hat{q}_\pm^2 = 4 \pm 2(\cos (q_x) + \cos (q_y) ).
\end{equation}
For antiferromagnetic correlations the parameters $\hat{q}^2$ and $\eta$ are 
\begin{eqnarray}
\hat{q}^2 & = & \hat{q}^2_+ \\
\eta(\hat{q}) & = & T_{sf} \hat{q}_-,
\end{eqnarray}
where $T_{sf}$ is a characteristic spin-fluctuation temperature. In the 
ferromagnetic case
\begin{eqnarray}
\hat{q}^2 & = & \hat{q}^2_- \\
\eta(\hat{q}) & = & T_{sf} \hat{q}_-.
\end{eqnarray}

We obtain the total spin-fluctuation propagator on the imaginary axis
$\chi({\bf q},i\nu_n)$ via the spectral representation
\begin{equation}
\chi({\bf q},i\nu_n) = - \int_{-\infty}^{+\infty} 	
	\frac{\mathrm{d}\omega}
	{\pi} \frac{\mbox{Im}\chi({\bf q},\omega)}
	{i\nu_n-\omega} .
\label{eq:chi}
\end{equation}
To get $\chi({\bf q},i\nu_n)$ to decay as $1/\nu_n^2$ as
$\nu_n \rightarrow \infty$, as it should, we introduce a cutoff 
$\omega_{cut}$ and take $\mbox{Im}\chi({\bf q},\omega) = 0$ for $\omega 
\geq \omega_{cut}$. A natural choice for the cutoff is $\omega_{cut} 
= \eta(\widehat{q})\kappa_0^2$. 

Using the effective action in Eq.~(\ref{eq:effac}) and the dynamical 
susceptibility in Eq.~(\ref{eq:chi}), the two-dimensional mean-field 
Eliashberg equations for the transition temperature $T_c$ in the 
Matsubara representation reduce to

\begin{eqnarray}
\Sigma({\bf p},i\omega_n) & = & g^2 \frac{T}{N} \sum_{\Omega_n}
	\sum_{\bf k} \chi({\bf p-k},i\omega_n-i\Omega_n)
	G({\bf k},i\Omega_n) 
\label{eq:sigma} 
\\
G({\bf p},i\omega_n) & = & \frac{1}
	{i\omega_n-(\epsilon_{\bf p}-\mu)-\Sigma({\bf p},i\omega_n)}  \\
\lambda(T) \Phi({\bf p},i\omega_n) & = & 
\left[
\begin{array}{c}
g^2/3 \\
-g^2
\end{array}
\right]
	\frac{T}{N} \sum_{\Omega_n} \sum_{\bf k} 
	\chi({\bf p-k},i\omega_n-i\Omega_n) \nonumber \\
	& & \mbox{} \times |G({\bf k},i\Omega_n)|^2 
	\Phi({\bf k},i\omega_n), 
\label{eq:lambda}
\end{eqnarray}
where 
\begin{equation}
\lambda(T) = 1 \Rightarrow T = T_c.
\label{eq:condTc}
\end{equation}
$\Sigma({\bf p},i\omega_n)$, $G({\bf p},i\omega_n)$ and 
$\Phi({\bf p},i\omega_n)$ are the Fourier components of the 
quasiparticle self-energy, the one-particle Green's function and the 
anomalous self-energy respectively. In \mbox{Eq. (\ref{eq:lambda})}
the prefactor $-g^2$ is for singlet pairing while the prefactor 
$g^2/3$ is for triplet pairing.
 
We find an instability to a d-wave gap function with 
$\Phi({\bf p},i\omega_n)$ transforming as $\cos(p_x)-\cos(p_y)$ 
in the nearly antiferromagnetic case and an instability to a p-wave 
gap function with $\Phi({\bf p},i\omega_n)$ transforming as 
$\sin(p_x)$ or $\sin(p_y)$ in the nearly ferromagnetic case.

The momentum convolutions in Eqs.~(\ref{eq:sigma}) and~(\ref{eq:lambda})
were evaluated with the aid of a fast-Fourier-transform algorithm on a 
$128\times128$ lattice. The corresponding frequency sums were carried out 
using the renormalisation group technique of Pao and Bickers \cite{rgasc}.
Between $8$ and $16$ Matsubara frequencies are kept at each stage of the
renormalization group procedure. The renormalization procedure is started 
at a temperature $T_0=0.4t$ and the frequency sum cut-off used is 
$\Omega_c \approx 20t$. The renormalization procedure restricts us to 
discrete temperatures so that the point at which the condition in 
\mbox{Eq. (\ref{eq:condTc})} is met must be determined by interpolation. 
The discrete temperatures were sufficiently close that a linear 
interpolation was adequate. The renormalization procedure afforded us 
considerable savings in computer time and storage requirements. Because 
of this we were able to carry out a thorough analysis of the dependence 
of our results on the spin-fluctuation coupling parameter $g^2\chi_0/t$ 
and the inverse correlation length parameter $\kappa^2$.

To investigate how strongly the transition temperature is influenced 
by various frequency and momentum parts of the paramagnon spectral 
function, we add an infinitesimal amount of spectral weight at 
frequencies $\omega_0 > 0$ and $-\omega_0$ and wavevector ${\bf q}^*$ 
and numerically calculate the change in $T_c$. More specifically, the
paramagnon spectral weight is changed from $\mbox{Im} \chi({\bf q},\omega)$ to 
$\mbox{Im} (\chi({\bf q},\omega) + \delta \chi({\bf q},\omega))$ with

\begin{equation}
   \mbox{Im} \delta \chi({\bf q},\omega) = 
	\eta \pi 
	\{ \delta(\omega-\omega_0) - \delta(\omega+\omega_0) \}
	\frac{N}{N_{{\bf q}^*}}
	\sum_{{\bf q}^*_i} \delta_{{\bf q},{\bf q}^*_i} ,
\end{equation}

\noindent where $\eta$ is a positive infinitesimal dimensionless parameter. 
The sum over ${\bf q}^*_i$ includes ${\bf q}^*$ and those wavevectors 
in the Brillouin zone related to it by the symmetry operations of 
the lattice. $N_{{\bf q}^*}$ is the number of such wavevectors 
including ${\bf q}^*$ itself.  $N$ is the total number of allowed 
wavevectors in the Brillouin zone. 

In the Matsubara representation, the addition of this infinitesimal amount
of spectral weight corresponds to changing the effective interaction 
$V_{eff}({\bf q},i\nu_n) = g^2\chi({\bf q},i\nu_n)$ in Eqs.~(\ref{eq:sigma})
and~(\ref{eq:lambda}) to $V_{eff} + \delta V_{eff}$ where

\begin{equation}
\delta V_{eff}({\bf q},i\nu_n)  = 
	 \frac{\eta}{\chi_0} g^2\chi_0
	 \frac{2\omega_0}{\omega_0^2 + \nu_n^2} 
	 \frac{N}{N_{{\bf q}^*}} 
	 \sum_{{\bf q}^*_i} \delta_{{\bf q},{\bf q}^*_i},
\end{equation}

The total effective interaction thus depends on the parameters 
$\kappa$, $\kappa_0$, $g^2\chi_0/t$ as well as the ratio
$\eta/\chi_0$, and therefore so does $T_c$. For small $\eta$, 
one has 

\begin{equation}
T_c(\eta/\chi_0,{\bf q}^*,\omega_0,\dots) 
= T_c(0,{\bf q}^*,\omega_0,\dots) + \frac{\eta}{\chi_0}
\frac{dT_c}{d(\eta/\chi_0)}(0,{\bf q}^*,\omega_0,\dots)
\end{equation}

\noindent where $\dots$ denote all the other parameters in the problem.
The quantity 

\begin{equation}
\Delta T_c ({\bf q}^*,\omega_0) = 
\frac{\chi_0}{T_c}\frac{\mathrm{d}T_c}
{\mathrm{d}\eta}({\bf q}^*,\omega_0)\Bigg|_{\eta=0}
\label{eq:DeltaTc}
\end{equation}

\noindent is a measure of how sensitive $T_c$ is to an infinitesimal 
change in the value of Im$\chi$ at $({\bf q}^*,\omega_0)$. We calculate
this derivative using the finite-difference estimate 

\begin{equation}
\frac{\mathrm{d}T_c}
{\mathrm{d}(\eta/\chi_0)}({\bf q}^*,\omega_0)\Big|_{\eta=0} 
\approx 
\frac{T_c(\eta/\chi_0,{\bf q}^*,\omega_0) - 
T_c(0,{\bf q}^*,\omega_0)}{\eta/\chi_0}.
\end{equation}

The value of the parameter $\eta/\chi_0$ must be chosen small enough, 
such that the function $T_c(\eta/\chi_0)$ is approximately 
linear in the vicinity of $\eta/\chi_0$. But if $\eta/\chi_0$ is chosen
too small, then it becomes very difficult to obtain reliable numerical
estimates of the differences in $T_c$. We found that 
$\eta/\chi_0 = 4\times 10^{-4}\mbox{ }t$ was a good compromise in the 
nearly antiferromagnetic case. A smaller value of $\eta/\chi_0$ was 
necessary for some of the nearly ferromagnetic results.
In particular, we carried out the nearly ferromagnetic calculations 
for $\kappa^2 = 0.5$ and $g^2\chi_0/t = 10, 5$ and for 
$g^2\chi_0/t = 30$ and $\kappa^2 = 2, 3, 4$ with 
$\eta/\chi_0 = 5 \times 10^{-5}\mbox{ }t$. We estimate that the 
accuracy with which we calculated the derivative $\mathrm{d}T_c/\mathrm{d}\eta$ in 
all cases is of the order of a few per cent.

As in Reference~\cite{MillisSachdevVarma88}, we define a crossover frequency 
$\omega_{cross}({\bf q}^*)$ by $\Delta T_c({\bf q}^*,\omega_{cross}) = 0$. 

Similarly, one can define an optimal frequency $\omega_{opt} ({\bf q}^*)$
as the frequency where $\Delta T_c$ is maximum. $\omega_{opt}({\bf q}^*)$
then indicates which paramagnon frequency and wavevector ${\bf q}^*$
contribute most to pairing. 

In order to make a comparison with the corresponding electron-phonon 
problem it is instructive to define a mass renormalization 
parameter $\lambda_Z$ and interaction parameter $\lambda_\Delta$. We define

\begin{eqnarray}
\lambda_Z & = & { \int_{-\infty}^{+\infty} {d\omega\over \pi}
<{1\over \omega}\mbox{Im}V_Z({\bf p}-{\bf p'},\omega)>_{FS({\bf p},{\bf p'})} 
\over <1>_{FS({\bf p})} } \label{lambda1} \\
\lambda_\Delta & = & -{ \int_{-\infty}^{+\infty} {d\omega\over \pi}
<{1\over \omega}\mbox{Im}V_\Delta({\bf p}-{\bf p'},\omega)
\eta({\bf p})\eta({\bf p'})>_{FS({\bf p},{\bf p'})} 
\over <\eta^2({\bf p})>_{FS({\bf p})} } \label{lambda2}
\end{eqnarray}

\noindent where

\begin{equation}
V_Z({\bf q},\omega) = g^2\chi({\bf q},\omega)
\label{Vz}
\end{equation}

\noindent and

\begin{eqnarray}
V_p({\bf q},\omega) & = & -{g^2\over 3}\chi({\bf q},\omega) \\
\eta({\bf p}) & = & \sin(p_x)
\label{Vp}
\end{eqnarray}

\noindent for p-wave spin-triplet pairing $(\Delta \equiv p)$ while

\begin{eqnarray}
V_d({\bf q},\omega) & = & g^2\chi({\bf q},\omega) \\
\eta({\bf p}) & = & \cos(p_x) - \cos(p_y)
\label{Vd}
\end{eqnarray}

\noindent in the case of d-wave spin-singlet pairing $(\Delta \equiv d)$.
In carrying out the frequency integrations, we omitted 
the cutoff which was used in Eq.~(\ref{eq:chi}). 
The approximate effect of the cutoff is to multiply $\lambda_\Delta$ 
and $\lambda_Z$ by a common factor which is weakly dependent on 
wavevector and $\kappa^2$. This can be ignored since 
we will only be interested in the quotient $\lambda_\Delta / \lambda_Z$.
The Fermi-surface averages are given by

\begin{eqnarray}
<\cdots>_{FS({\bf p})}  & = & \int {d^2p\over (2\pi)^2} \cdots 
\delta(\epsilon_{\bf p} - \mu) \label{FSaverage1} \\
<\cdots>_{FS({\bf p},{\bf p'})}  & = & \int {d^2p\over (2\pi)^2} 
{d^2p'\over (2\pi)^2}\cdots \delta(\epsilon_{\bf p} - \mu) 
\delta(\epsilon_{\bf p'} - \mu)  \label{FSaverage2}
\end{eqnarray}

\noindent In practice, we compute the Fermi-surface average 
with a discrete set of wavevectors on a lattice and we replace the 
delta function by a finite-temperature expression

\begin{eqnarray}
\int {d^2p\over (2\pi)^2} & \longrightarrow &{1\over N}\sum_{\bf p} \\
\delta(\epsilon_{\bf p} - \mu) & \longrightarrow & {1\over T} 
f_{\bf p}(1-f_{\bf p})
\label{FSaverageNum}
\end{eqnarray}

\noindent where $f_{\bf p}$ is the Fermi function and $N$ is the number
of wavevectors in The Brillouin Zone.
Note that ${1\over T} f_{\bf p}(1-f_{\bf p}) 
\rightarrow \delta(\epsilon_{\bf p} - \mu)$ as $T 
\rightarrow 0$. We used $T = 0.1t$ and $N = 128^2$.
The finite temperature effectively means that 
van Hove singularities will be smeared out.
The Fermi-surface average that appears in $\lambda_Z$, 
Eq.~(\ref{lambda1}), plays a role similar to that of 
$\alpha^2F(\omega)/\omega$ in the case of phonon-mediated 
superconductivity. 

\section{Results}
\noindent
The model consists of the parameters $g^2\chi_0/t$, 
$T_{sf}/t$, $\kappa_0$ and $\kappa$. It is found experimentally 
that $T_{sf}\kappa^2_0$ is approximately constant. We shall use this 
relation to eliminate one parameter from the set and pick a 
representative value of the product $T_{sf}\kappa^2_0$.
We put $T_{sf} = \frac{2}{3}t$ which corresponds to a temperature 
of $1000$~K if the bandwidth is $1$~eV, and $\kappa_0^2a^2=12$ for 
comparison with earlier work carried out on this model\cite{pvsdMon99}.
To obtain a representative value for the dimensionless coupling parameter 
$g^2\chi_0/t$, we make use of the Stoner criterion. In the vicinity
of the magnetic instability, $g \chi_0({\bf Q},0) \approx 1$, where
$\chi_0({\bf q},\omega)$ is the usual tight-binding Lindhard 
susceptibility and ${\bf Q}$ the ordering vector. With 
$\chi_0({\bf Q},0)\approx N(0) \approx \frac{1}{8t}$ where $N(0)$ is the 
single-particle density of states at the Fermi level, the value of the 
coupling parameter $g^2\chi_0/t$ obtained is about 10. We stress that 
this is only an order of magnitude estimate of the coupling parameter.

Figures~1 and 2 show
$\Delta T_c ({\bf q}^*,\omega_0)$, Eq.(\ref{eq:DeltaTc}),
versus frequency $\omega_0$ for several values of (a) 
$g^2\chi_0/t$, (b) $\kappa ^2$ and (c) $\bf q^*$. 
Figure~1 shows the results in the case of a nearly antiferromagnetic 
metal and Figure~2 shows the results in the case of a 
nearly ferromagnetic metal. By contrast to the phonon case
where $\Delta T_c$ is always positive, note that there are wavevectors
${\bf q}^*$ for which $\Delta T_c$ is always negative, which means
that spectral weight at these wavevectors is deleterious to 
superconductivity. And if spectral weight is added at sufficiently 
small frequencies $\omega_0$, $\Delta T_c$ is negative for all 
wavevectors ${\bf q}^*$. Finally, note that the overall scale of
$\Delta T_c$ for nearly ferromagnetic systems is larger than that for
nearly antiferromagnetic systems. This indicates a larger sensitivity
of the relative changes in $T_c$ with respect to changes in the
spin-fluctuation spectral weight for nearly ferromagnetic systems.

Figure~3 is a representative plot of 
\begin{equation}
\Delta T_c^+ = \mbox{max}\left\{\frac{\chi_0}{T_c}\frac{\mathrm{d}T_c}
{\mathrm{d}\eta},0\right\}
\end{equation}
as a function of the wavevector ${\bf q}^*$ at which
spectral weight is added for several values of $\omega_0/t$ and 
for a nearly antiferromagnetic metal.
$\Delta T_c^+$ indicates the region of momentum and frequency space
where adding a small amount of spectral weight enhances the 
superconducting transition temperature.
Figure~4 is a complementary plot for
\begin{equation}
\Delta T_c^- = \mbox{min}\left\{\frac{\chi_0}{T_c}\frac{\mathrm{d}T_c}
{\mathrm{d}\eta},0\right\},
\end{equation}
which indicates the region of momentum and frequency space
where adding a small amount of spectral weight suppresses
the superconducting transition temperature.
These results should be compared with 
the corresponding plots, shown in Figures~5 ($\Delta T_c^+$) 
and 6 ($\Delta T_c^-$), for the nearly ferromagnetic case. 
We have chosen the values of $\omega_0$ for graphs (a-c) to be less than
$\omega_{opt}$ (which, for these parameter values, 
is about $2.5t$ in the nearly antiferromagnetic case and about 
$0.32t$ in the nearly ferromagnetic case) and the values of $\omega_0$ in 
graph (d)
to be greater than $\omega_{opt}$. The
region in wavevector space where $\Delta T_c \geq 0$ grows 
monotonically with $\omega_0$, starting from zero when 
$\omega_0 < \omega_{cross}$, the crossover frequency corresponding to the
incipient ordering wavevector. However the maximum value of $\Delta T_c$ passes
through an extremum at $\omega_0 = \omega_{opt}$, the optimal frequency
corresponding to the incipient ordering wavevector.  
It is interesting to find that the size of the region  $\Delta T_c \geq 0$ 
is approximately the same in the nearly antiferromagnetic and
nearly ferromagnetic cases if $\omega_0$ is chosen close
to the corresponding optimal frequency $\omega_{opt}$, and that it
is a significant fraction of the total area of the Brillouin Zone.
This suggests that except for very small values of $\kappa^2$, 
the pairing occurs mostly through the short-ranged magnetic fluctuations. 
Comparing the values of $\Delta T_c$ in the frequency and
wavevector scans suggests that removing spectral weight from 
low frequencies may be an effective way to enhance $T_c$.

We show in Figure~7 plots of $\Delta T_c^+$ for various values of
$\kappa^2$ in the nearly ferromagnetic case. The surfaces are very 
similar to one another in the small $\kappa^2$ limit. This can also be 
seen in the $\Delta T_c$ versus $\omega_0$ curves. However, since 
our mean-field calculations are likely not accurate very close to 
the critical point for magnetic order, we cannot state categorically 
what the limiting values of $\Delta T_c$ are. The $\Delta T_c = 0$ 
contour in wavevector space is quite insensitive to the value 
of $\kappa^2$, but is, however, very sensitive to the value of $\omega_0$. 

Figure~8 shows how the crossover frequency $\omega_{cross}({\bf q}^*)$
depends on the coupling constant $g^2\chi_0/t$ and the correlation
wavevector $\kappa^2$, for the optimum wavevector, namely 
${\bf q}^* = (\pi,\pi)$ for nearly antiferromagnetic systems and
${\bf q}^* = (0,0)$ for nearly ferromagnetic systems.
The figure shows that $\omega_{cross}({\bf q}^*)$ when scaled by $T_c$,
is more robust to changes in $g^2\chi_0/t$ and $\kappa^2$ in 
the nearly antiferromagnetic case than in the nearly ferromagnetic
case. The value of $\omega_{cross}$ in the nearly antiferromagnetic case 
is about $3T_c$ which is quite close to the value $2T_c$ obtained 
in the Hubbard model\cite{ms94}. The value of $\omega_{cross}$ 
in the nearly ferromagnetic case lies between $5T_c$ and $9T_c$.
It is interesting to observe that in the limit of small coupling 
and far away from the magnetic instability, $\omega_{cross}/T_c$ 
seems to approach a value common to both the nearly antiferromagnetic 
and nearly ferromagnetic cases. Figure~8(a) indicates that, 
for $\kappa^2=0.50$ in the nearly ferromagnetic case,
$\omega_{cross}/T_c$ seems to depend on $T_{sf}$ and $\kappa_0^2$ 
only through the product $T_{sf}\kappa_0^2$, which is found 
experimentally to be approximately constant.
Therefore $\omega_{cross}$ is roughly proportional to $T_c$.
But, for the same parameters, $T_c$ scales approximately linearly 
with $T_{sf}$\cite{pvsdMon99}. Hence, for $\kappa^2=0.50$ 
in the nearly ferromagnetic case, $\omega_{cross}$ scales approximately 
linearly with $T_{sf}$.   

One would like to understand the nature of the crossover frequency
$\omega_{cross}$ in terms of phonon-problem-like parameters.
Millis and coworkers\cite{MillisSachdevVarma88} 
showed that, in the antiferromagnetic magnetic fluctuation case,

\begin{equation}
\omega_{cross}/T_c \sim e^\frac{1}{\gamma_d},
\label{omegaCrossD}
\end{equation} 

\noindent where $\gamma_d$ is a d-wave effective interaction constant 
$\gamma_d = \frac{\lambda_d}{\lambda_Z}$. 
They assumed that the dynamical magnetic susceptibility can be seperated
into a wavevector-dependent factor and a frequency-dependent factor and 
they adopted an Einstein model to describe the frequency-dependent 
factor. They further assumed that 
the \'{E}liashberg equations can be reduced to equations 
involving Fermi-surface quantities only. 
A similar calculation can be carried out in the nearly ferromagnetic case 
yielding

\begin{equation}
\omega_{cross}/T_c \sim e^\frac{1}{\gamma_p},
\label{omegaCrossP}
\end{equation} 

\noindent with $\gamma_p = \frac{\lambda_p}{\lambda_Z}$.
The results of the more complete calculation described in Section II 
are compared with these expressions for $\omega_{cross}/T_c$ in Figure~9. 
Since $\gamma_\Delta$ ($\Delta=p,d$) is not an independent 
parameter in our model, we calculated
$\log (\omega_{cross}/T_c)$ and 
$1/\gamma_\Delta$ for various values of $\kappa^2$. 
One sees that the expressions given in 
Eqs.~(\ref{omegaCrossD}) and~(\ref{omegaCrossP}) 
are a poor approximation to the value of $\omega_{cross}/T_c$
calculated from the Eliashberg equations.
The dependence of $\omega_{cross}/T_c$ on $\gamma_\Delta$ is 
much weaker than $e^\frac{1}{\gamma_\Delta}$ and is not even monotonic 
in the nearly antiferromagnetic case. This finding is consistent 
with the inability to obtain a simple analytic expression similar 
to that proposed by McMillan to represent the $T_c$ calculated 
numerically via the \'{E}liashberg equations\cite{pvsdMon99,3DMon01}.

We show in Figure~10 the variation of the normalised
optimal frequency $\omega_{opt}({\bf q}^*)/T_c$ with
$g^2\chi_0/t$ and $\kappa^2$ in the nearly antiferromagnetic case,  
with ${\bf q^*} = (\pi,\pi)$, and nearly ferromagnetic case 
with ${\bf q^*} = (0,0)$. In the nearly antiferromagnetic case 
$\omega_{opt}$ is approximately a constant equal to $10T_c$,
which is close to the value it takes in the Hubbard model\cite{ms94}.
It is also of the same order of magnitude as $2\pi T_c$, the value 
it takes in the electron-phonon problem\cite{br73}.
In the nearly ferromagnetic case $\omega_{opt}$ lies between $13T_c$ 
and $22T_c$.

As shown by Bergmann and Rainer for the phonon problem\cite{br73}, 
it is instructive to compare the frequency dependence of the spectral function 
with that of $\Delta T_c$. The results for the nearly antiferromagnetic case
and ${\bf q^*} = (\pi,\pi)$ are shown in Figure~11 for various values of 
$\kappa^2$.

Figure~11 shows that in the small $\kappa^2$ limit (strong coupling)
the peak in the spectral weight lies below the optimum frequency 
$\omega_{opt}$, while for large $\kappa^2$ (weak coupling)
it lies above $\omega_{opt}$.
This is analogous to the results of Bergmann and Rainer who found that 
the transverse phonon mode was below the optimum frequency 
$\omega_{opt} = 2\pi T_c$ for strong
coupling superconductors such as Hg and above $\omega_{opt}$
for weak coupling superconductors such as In. 
The results of Figure~11 suggest an explanation for a maximum $T_c$ as a function
of $\kappa^2$. As $\kappa^2\rightarrow 0$, while the d-wave component of
the pairing interaction increases, the frequency at which
the spin-fluctuation spectral weight is maximum becomes smaller than
$\omega_{cross}$, the frequency below which addition
of spectral weight produces a suppression of $T_c$.
 
One might have expected that the $\kappa^2$ for which
$T_c$ is maximum would be such that the peak in the spin-fluctuation spectral
weight coincides with $\omega_{opt}$. But for $g^2\chi_0/t = 10$, the maximum
$T_c$ is found at $\kappa^2 \approx 0.35$, while the match between the peak
in the spectral weight and $\omega_{opt}$ occurs for $\kappa^2 \approx 1$.
The shift in the peak of Im$V_d$ relative to $\omega_{opt}$ allows one to
understand the trends in $T_c$, but the arguments remain qualitative.

Since the location of the peak in the spectral function does not depend
on the value of $g^2\chi_0/t$ and our results show that the frequency
$\omega_{opt}$ is approximately constant in the nearly antiferromagnetic case,
one does not gain new insights by looking at the frequency dependences
of Im$V_d({\bf q^*},\omega_0)$ and $\Delta T_c({\bf q^*},\omega_0)$ 
for different coupling constants.

In the nearly ferromagnetic case, since Im$\chi({\bf q}\rightarrow 0,\omega) = 0$
for $\omega \neq 0$, one sees that the situation is quite different and thus
the shape of the curves Im$V_p({\bf q^*},\omega_0)$ and 
$\Delta T_c({\bf q^*},\omega_0)$ for the optimum wavevector ${\bf q^*} = 0$
will not be similar, unlike the phonon and 
nearly antiferromagnetic cases.
\section{Discussion}
\noindent Our results demonstrate a number of significant differences
as well as similarities between the magnetic fluctuation- and
phonon-pairing mechanisms. In the latter case, the interaction 
is local in space but non-local in time. This means that there should 
be almost no variation of $\Delta T_c({\bf q^*},\omega_0)$ with 
${\bf q^*}$ and therefore one only need worry about the dependence 
of $\Delta T_c$ on frequency. In the case of magnetic pairing however,
the interaction is non-local in both space and time and as a result
$\Delta T_c({\bf q^*},\omega_0)$ exhibits a dependence on both 
wavevector ${\bf q^*}$ and frequency $\omega_0$.

In the phonon case\cite{br73}, addition of an infinitesimal amount of spectral 
weight at any non-zero frequency $\omega_0$ results in an enhancement 
of $T_c$, with an optimum frequency of $2\pi T_c$. In the limit
$\omega_0\rightarrow 0$, $\Delta T_c\rightarrow 0$, which is consistent
with Anderson's theorem\cite{Anderson} for isotropic superconductors.
On the other hand, for the magnetic interaction model, addition of spectral
weight at sufficiently low frequencies results in negative values of 
$\Delta T_c$. 
The intuitive understanding of this result\cite{MillisSachdevVarma88} is also
related to Anderson's theorem. Very low frequency paramagnons effectively 
act as a static non-magnetic
impurity potential that scatters the quasiparticles. For anisotropic 
superconductors, Anderson's theorem does not apply and the presence
of non-magnetic impurities leads to a supression of $T_c$. Also note
that for certain wavevectors ${\bf q^*}$, $\Delta T_c({\bf q^*},\omega_0)$ 
is always negative, regardless of $\omega_0$.

There are also significant differences between the nearly antiferromagnetic 
and nearly ferromagnetic cases. For instance, the crossover frequency
$\omega_{cross}$ at which $\Delta T_c({\bf q^*},\omega_0) = 0$
depends more
weakly on the model parameters $\kappa^2$ and $g^2\chi_0/t$ in the
nearly antiferromagnetic case than in the nearly ferromagnetic case.
($\omega_{cross}$ is 
strongly dependent on ${\bf q^*}$ in both
cases.)
Similarly, the optimum frequency  $\omega_{opt}$ at which 
$\Delta T_c({\bf q^*},\omega_0)$ is maximum is more weakly dependent on the 
model parameters $\kappa^2$ and $g^2\chi_0/t$ in the nearly antiferromagnetic 
metals than in the nearly ferromagnetic metals.
Moreover, the overall scale of $\Delta T_c$ is much larger in the former case 
than in the latter, indicating a much greater relative sensitivity 
of $T_c$ to changes in the spectral weight for nearly ferromagnetic systems.

Our results also show similarities between magnetically mediated 
and phonon-mediated superconductivity. In all cases, the shape
of the $\Delta T_c$ curves are monotonic with a single optimum 
frequency (for each wavevector) and, in the high-frequency regime 
and for wavevectors such that $\Delta T_c >0$, the curves
behave similarly . The decrease of $\Delta T_c$ with increasing 
$\omega_0$ approximately mirrors the change in the interaction 
parameters $\lambda_\Delta$ as $V_{eff}\rightarrow V_{eff}+\delta V_{eff}$.
The relative positions of the peak $\omega_{max}$ in the spectral weight 
Im$V_d({\bf q^*}=(\pi,\pi),\omega)$ and optimum frequency $\omega_{opt}$ in
the strong coupling limit where $\omega_{max} < \omega_{opt}$ and weak coupling
limit in which $\omega_{max} > \omega_{opt}$ is similar in some respects 
to the findings of Bergmann and Rainer\cite{br73} for the phonon problem. 
In that case, the transverse phonon frequency is below the optimum one for 
strong coupling superconductors and above in the weak coupling case.

Figure 2(c) shows a feature that may partially explain the difficulty in
observing superconductivity on the border of itinerant ferromagnetism.
Our results (see Figure 2(c)) show that adding spectral weight at large 
wavevectors is detrimental to $T_c$. In other words, antiferromagnetic 
correlations act to suppress pairing in nearly ferromagnetic systems.
In the presence of a lattice, one would generically expect some 
enhancement of the magnetic response at large wavevectors due to
nesting features in the Fermi surface for nearly-half-filled electronic
bands. As explained in References ~\cite{pvsdMon99,3DMon01}, magnetic 
pairing on the border of long-range ferromagnetic order is not as robust as
magnetic pairing in nearly antiferromagnetic systems. The antiferromagnetic 
correlations inherent to the presence of a crystal lattice tend to suppress 
magnetic pairing in nearly ferromagnetic systems, thus making it even 
more difficult to observe.
\section{Outlook}
\noindent
The dependence of the superconducting transition temperature on model
parameters and the role played by dimensionality for Ornstein-Zernike like
spin-fluctuation spectra has been studied in some detail\cite{pvsdMon99,3DMon01}.
However, these calculations reflect an integral property of the dynamical
susceptibility $\chi({\bf q},\omega)$. In this paper, we presented a study of
how different regions in wavevector ${\bf q}$ and frequency $\omega$ individually
contribute to $T_c$, giving us novel insights into the magnetic interaction
model.

The calculations were carried out by adding infinitesimal amounts of 
spectral weight at different wavevectors and frequency and determining
the resulting changes in $T_c$ from the numerical solution of the
Eliashberg equations. We stress that the work reported here provides insight 
only into how different regions in ${\bf q}$ and $\omega$ contribute to 
pairing for the particular magnetic-fluctuation spectrum $\chi({\bf q},\omega)$, 
Eq.~(\ref{eq:chi}). One may infer certain trends in $T_c$ for small changes 
in the spin-fluctuation spectrum, but it may not be warranted to extrapolate 
our results to large changes in the magnetic spectrum. Also, the calculations
are carried out at $T_c$. They may give one some idea of how the robustness of
pairing is affected as one goes below $T_c$ and the electronic gap induces
changes in the momentum and frequency structure of $\chi({\bf q},\omega)$, but
this involes another extrapolation to large changes in the magnetic spectrum
which may not be warranted.

We found that addition of spectral weight at or near the incipient ordering 
wavevector in the nearly antiferromagnetic and nearly ferromagnetic cases
lead to an enhancement of $T_c$, except at low frequencies where it leads
to a suppression of the superconducting transition temperature. However, 
addition of spectral weight far away from the incipient ordering 
wavevector results in a lowering of $T_c$ regardless of frequency.
These results are in stark contrast to those obtained for phonon-mediated
superconductivity where addition of spectral weight at any non-zero frequency
leads to an enhancement of the superconducting critical temperature.

The theoretical framework presented here to describe systems on the border of 
magnetism can be translated to describe systems on the border of other types 
of instabilities, such as charge-density-wave or ferroelectric instabilities.
The same type of calculations could be carried out in those other cases, and
compared to the results of this paper, shedding light on the similarities 
and differences between the many possible pairing mechanisms.
\section{Acknowledgments}
\noindent
We acknowledge the support of the Isaac Newton Trust, the Cambridge
European Trust, the Robert Gardiner Memorial Scholarship and the 
EPSRC. P. McHale acknowledges the hospitality of the MPIPKS, Dresden where the
text was completed. 
Finally we thank G.G. Lonzarich for discussions on these
and related topics.
\begin{figure}
\caption{
Frequency dependence of $\Delta T_c = \frac{\chi_0}{T_c}\frac{\mathrm{d}T_c}
{\mathrm{d}\eta}$ 
for various values of 
(a) the coupling parameter $g^2\chi_0/t$, (b) the inverse correlation 
length $\kappa^2$ and (c) the momentum transfer ${\bf q}^*$ 
in the case of a nearly antiferromagnetic metal with a $d_{x^2-y^2}$ 
superconducting state symmetry. The characteristic spin-fluctuation 
temperature is $T_{sf} = 0.67$ and $\kappa_0^2 = 12$. 
}
\label{fig1}
\end{figure}
\begin{figure}
\caption{
Frequency dependence of $\Delta T_c = \frac{\chi_0}{T_c}\frac{\mathrm{d}T_c}
{\mathrm{d}\eta}$ 
for various values of 
(a) the coupling parameter $g^2\chi_0/t$, (b) the inverse correlation 
length $\kappa^2$ and (c) the momentum transfer ${\bf q}^*$ 
in the case of a nearly ferromagnetic metal with a p-wave
superconducting state symmetry. The characteristic spin-fluctuation 
temperature is $T_{sf} = 0.67$ and $\kappa_0^2 = 12$. 
}
\label{fig2}
\end{figure}
\begin{figure}
\caption{
Wavenumber dependence of 
$ \Delta T^+_c = \mbox{min}\{\frac{\chi_0}{T_c}\frac{\mathrm{d}T_c}
{\mathrm{d}\eta},0\}$ 
for various values of $\omega_0/t$ for a nearly antiferromagnetic
metal with a $d_{x^2-y^2}$ superconducting state symmetry. 
The other pertinent parameter values are 
$g^2\chi_0/t = 30$ and $\kappa^2 = 0.5$.
}
\label{fig3}
\end{figure}
\begin{figure}
\caption{
Wavenumber dependence of 
$ \Delta T^-_c = \mbox{min}\{\frac{\chi_0}{T_c}\frac{\mathrm{d}T_c}
{\mathrm{d}\eta},0\}$ 
for various values of $\omega_0/t$ for a nearly antiferromagnetic
metal with a $d_{x^2-y^2}$ superconducting state symmetry. 
The other pertinent parameter values are 
$g^2\chi_0/t = 30$ and $\kappa^2 = 0.5$.
}
\label{fig4}
\end{figure}
\begin{figure}
\caption{
Wavenumber dependence of 
$ \Delta T^+_c = \mbox{max}\{\frac{\chi_0}{T_c}\frac{\mathrm{d}T_c}
{\mathrm{d}\eta},0\}$ 
for various values of $\omega_0/t$ for 
a nearly ferromagnetic metal with a p-wave
superconducting state symmetry.
The other pertinent parameter values are 
$g^2\chi_0/t = 30$ and $\kappa^2 = 0.5$.
}
\label{fig5}
\end{figure}
\begin{figure}
\caption{
Wavenumber dependence of 
$ \Delta T^-_c = \mbox{min}\{\frac{\chi_0}{T_c}\frac{\mathrm{d}T_c}
{\mathrm{d}\eta},0\}$ 
for various values of $\omega_0/t$ for
a nearly ferromagnetic metal with a p-wave
superconducting state symmetry. 
The other pertinent parameter values are 
$g^2\chi_0/t = 30$ and $\kappa^2 = 0.5$.
}
\label{fig6}
\end{figure}
\begin{figure}
\caption{
Wavenumber dependence of 
$ \Delta T^+_c = \mbox{max}\{\frac{\chi_0}{T_c}\frac{\mathrm{d}T_c}
{\mathrm{d}\eta},0\}$ 
for various values of the dimensionless parameter $\kappa^2$ for
a nearly ferromagnetic metal with a p-wave
superconducting state symmetry.
The other pertinent parameter values are 
$g^2\chi_0/t = 30$ and $\omega_0/t = 2$.
}
\label{fig7}
\end{figure}
\begin{figure}
\caption{
The dependence of the crossover frequency $\omega_{cross}$, normalised
by the superconducting transition temperature $T_c$, on (a) the coupling
parameter $g^2\chi_0/t$ and on (b) the inverse correlation length
$\kappa^2$. Both graphs exhibit results obtained in the nearly
antiferromagnetic (NA) case and for three values of the parameter pair 
$(T_{sf},\kappa^2_0)$, keeping the product $T_{sf}\kappa_0^2$ constant,
in the nearly ferromagnetic (NF) case. 
Throughout we have put 
${\bf q}^* = (\pi,\pi)$ in the nearly antiferromagnetic case
and ${\bf q}^* = (0,0)$ in the nearly ferromagnetic case. 
}
\label{fig8}
\end{figure}
\begin{figure}
\caption{Comparison of our results for the crossover frequency in (a) the
nearly antiferromagnetic case (circles) and 
(b) the nearly ferromagnetic case (circles) with the result
$\omega_{cross}/T_c \sim e^\frac{1}{\gamma_\Delta}$, Eq.~(\ref{omegaCrossD})
(squares). We have put
the prefactor in this formula equal to one for the sake of comparison.
We have calculated $\omega_{cross}$ with $g^2\chi_0/t = 10$, although
the effective interaction constant $\gamma_\Delta$ is independent 
of $g^2\chi_0/t$.
The data is parametrised left to right by increasing $\kappa^2$ 
from (a) $0.10$ to $4.00$ and (b) $0.10$ to $1.00$.
Throughout we have put 
${\bf q}^* = (\pi,\pi)$ in the nearly antiferromagnetic case
and ${\bf q}^* = (0,0)$ in the nearly ferromagnetic case. 
}
\label{fig9}
\end{figure}
\begin{figure}
\caption{The dependence of the optimal frequency $\omega_{opt}$, normalised
by the superconducting transition temperature $T_c$, on 
(a) $g^2\chi_0/t$ and (b) 
$\kappa^2$.
We have put $T_{sf} = 0.67$, $\kappa_0^2 = 12$ and ${\bf q}^*=(\pi,\pi)$
in the nearly antiferromagnetic case and ${\bf q}^*=(0,0)$ in the 
nearly ferromagnetic case.
}
\label{fig10}
\end{figure}
\begin{figure}
\caption{Comparison of $\Delta T_c = \frac{\chi_0}{T_c} \frac{dT_c}{d\eta}$ 
(solid line) and Im$V_d$ (dashed line) for various values of the parameter 
$\kappa^2$ for a nearly antiferromagnetic metal. The functions are plotted for a
value of the coupling $g^2\chi_0/t = 10$ and a wavevector
${\bf q}^* = (\pi,\pi)$.
}
\label{fig11}
\end{figure}
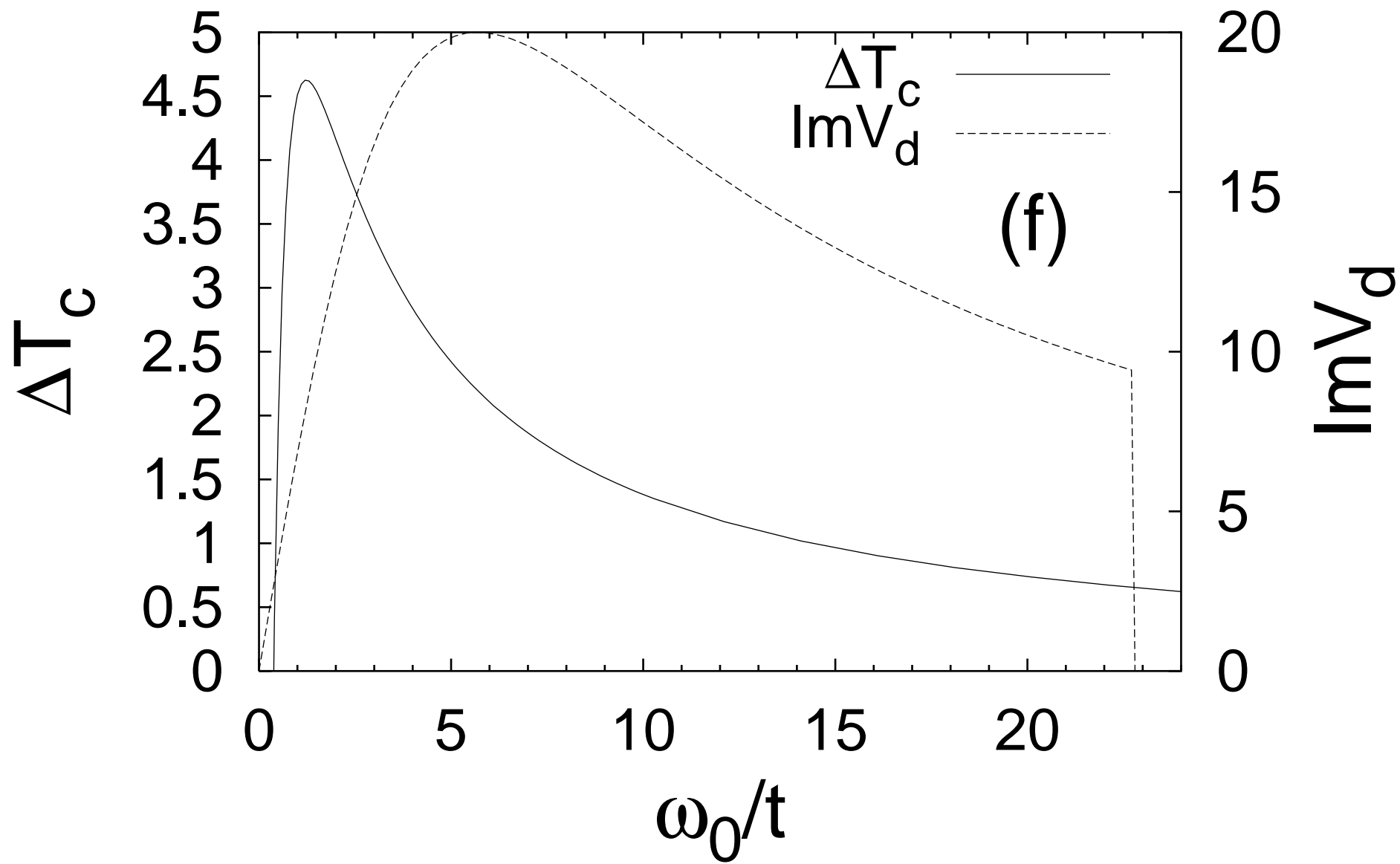
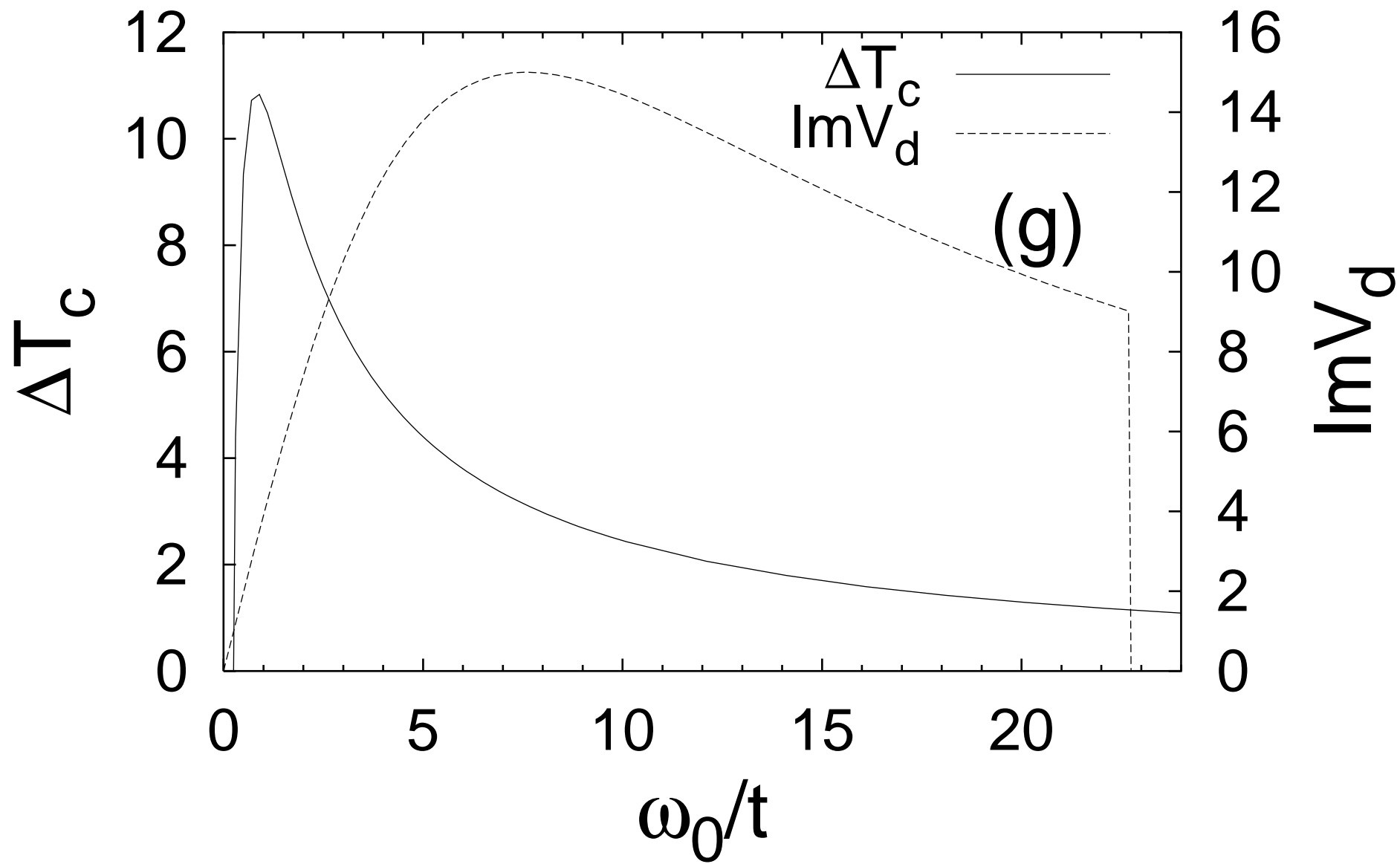
\end{document}